\documentclass[10pt,letterpaper]{article}
\usepackage[top=0.85in,left=2.75in,footskip=0.75in]{geometry}

\usepackage{amsmath,amssymb}

\usepackage{changepage}

\usepackage{textcomp,marvosym}

\usepackage{cite}

\usepackage{nameref,hyperref}

\usepackage[nopatch=eqnum]{microtype}
\DisableLigatures[f]{encoding = *, family = * }

\usepackage[table]{xcolor}

\usepackage{array}

\newcolumntype{+}{!{\vrule width 2pt}}

\newlength\savedwidth

\raggedright
\usepackage[aboveskip=1pt,labelfont=bf,labelsep=period,justification=raggedright,singlelinecheck=off]{caption}

\makeatletter
\renewcommand{\@biblabel}[1]{\quad#1.}
\makeatother

\usepackage{lastpage,fancyhdr,graphicx}
\usepackage{epstopdf}
\fancyheadoffset[L]{2.25in}
\fancyfootoffset[L]{2.25in}
\begin{document}
\vspace*{0.2in}

\begin{flushleft}
{\Large
\textbf\newline{Addressing overlapping communities in multiple-source detection: An edge clustering approach for complex networks} % Please use "sentence case" for title and headings (capitalize only the first word in a title (or heading), the first word in a subtitle (or subheading), and any proper nouns).
}
\newline
\\
Haomin Li\textsuperscript{1},
Daniel K. Sewell\textsuperscript{1*}
\\
\bigskip
\textbf{1} Department of Biostatistics, University of Iowa, Iowa City, IA, USA
\bigskip

% Insert additional author notes using the symbols described below. Insert symbol callouts after author names as necessary.
% 
% Remove or comment out the author notes below if they aren't used.
%
% Primary Equal Contribution Note
%\Yinyang These authors contributed equally to this work.

% Additional Equal Contribution Note
% Also use this double-dagger symbol for special authorship notes, such as senior authorship.
% \ddag These authors also contributed equally to this work.

%\textcurrency Current Address: Dept/Program/Center, Institution Name, City, State, Country % change symbol to "\textcurrency a" if more than one current address note
% \textcurrency b Insert second current address 
% \textcurrency c Insert third current address


% Group/Consortium Author Note
%\textpilcrow Membership list can be found in the Acknowledgments section.

% Use the asterisk to denote corresponding authorship and provide email address in note below.
* daniel-sewell@uiowa.edu

\end{flushleft}
\section*{Abstract}
The source detection problem in network analysis involves identifying the origins of diffusion processes, such as disease outbreaks or misinformation propagation. Traditional methods often focus on single sources, whereas real-world scenarios frequently involve multiple sources, complicating detection efforts. This study addresses the multiple-source detection (MSD) problem by integrating edge clustering algorithms into the community-based label propagation framework, effectively handling mixed-membership issues where nodes belong to multiple communities.

The proposed approach applies the automated latent space edge clustering model to a network, partitioning infected networks into edge-based clusters to identify multiple sources. Simulation studies on ADD HEALTH social network datasets demonstrate that this method achieves superior accuracy, as measured by the F1-Measure, compared to state-of-the-art clustering algorithms. The results highlight the robustness of edge clustering in accurately detecting sources, particularly in networks with complex and overlapping source regions. This work advances the applicability of clustering-based methods to MSD problems, offering improved accuracy and adaptability for real-world network analyses.

% Use "Eq" instead of "Equation" for equation citations.
\section*{Introduction}
\label{sec:intro}
The source detection problem is a critical topic in network analysis. It involves identifying the origin or source(s) responsible for diffusion processes, such as the outbreak of epidemics, the spread of gossip over online social networks, the propagation of computer viruses over the Internet, and the adoption of innovations~\cite{zhu2014information}. Identifying these sources is crucial because it allows for targeted interventions, which are often more effective and efficient than broader, less focused strategies, and leads to a better understanding of the diffusion process.

In epidemiology, for example, determining the initial carrier of a disease (often referred to as "patient zero") can inform precise containment strategies that prevent further transmission and save lives. Similarly, in cybersecurity, identifying the origin of a malware attack or computer virus enables rapid neutralization, minimizing damage and protecting vulnerable systems~\cite{antulov2015identification}.

Beyond healthcare and cybersecurity, source detection is essential in managing the spread of misinformation in social networks. Identifying the originator of false information allows platforms to intervene early, reducing the harmful impact on public opinion and social behavior. In marketing, understanding the key influencers or early adopters who drive the spread of new products or ideas can help optimize promotional strategies and accelerate diffusion~\cite{shah2010detecting}.

The effectiveness of source detection methods depends on the type of network observations utilized. In this study, we focused on snapshot-based observations. A snapshot provides a view of the network at a specific point in time, offering information about the nodes that are infected at the time of observation and the probability of infection for those nodes that have already been exposed to the diffusion process~\cite{zhu2014robust}.

Most source detection research assumes that there is a single source in the network when studying the diffusion process~\cite{shelke2019source}. However, in reality, the spread of misinformation or infections often originates from multiple sources rather than a single point. For example, during the early stages of the COVID-19 pandemic, the virus spread from multiple locations, as international and domestic travel facilitated the simultaneous introduction of the virus into a given locale. Similarly, diseases like Ebola and Zika often have multiple sources of outbreaks, especially when the disease is transmitted from animals to humans in different regions. These multiple introductions complicate the containment process and require detection methods that can identify more than one source of infection within a network of disease transmission ~\cite{holmdahl2020wrong, waldman2020epidemiology}.

Few researchers have focused on multiple sources detection (MSD) problem. One such method builds off of the Rumor Centrality, initially designed for single-source detection, in order to handle multiple sources by considering infection times and network topology, thereby allowing for identification of multiple origins in the diffusion process~\cite{shah2010detecting}. Another method, Backtracking Source Identification, works by reversing the infection spread, attempting to retrace the diffusion paths to pinpoint the potential origins of the spread~\cite{dong2013rooting}.

Zang et.al~\cite{zang2014discovering} first proposed a clustering-based method for the MSD problem. Their method involves first performing clustering on the infected network to partition it into distinct groups. Once the clustering structure is identified, a single-source detection algorithm is applied within each cluster to locate the sources independently ~\cite{zang2015locating}.

Simulation studies have shown that treating MSD as a series of single-source detection problems is insufficient. This limitation may arise in part from disregarding the influence and interactions between neighboring clusters. To overcome this, Zhang et.al~\cite{zhang2023clp} introduced a community-based label propagation (CLP) framework, building on the work of Wang et.al~\cite{wang2017multiple}. The CLP framework begins with clustering the infection network but improves detection accuracy by incorporating node prominence and exoneration effects, where nodes surrounded by more infected nodes and fewer non-infected nodes are more likely to be identified as sources. %\textcolor{red}{You mention two issues: (1) clustering errors, and (2) ignoring interactions between clusters, and that these stem from treating MSD as a set of SSD problems.  Does Zhang actually overcome these issues?  It's a bit more than a series of SSD problems, but do they address clustering errors and interactions between clusters?  Maybe the latter, but I'm not sure they (or your algorithm, for that matter), address the former.} 

\begin{figure}[!h]
 \centering
  \includegraphics[scale=0.45]{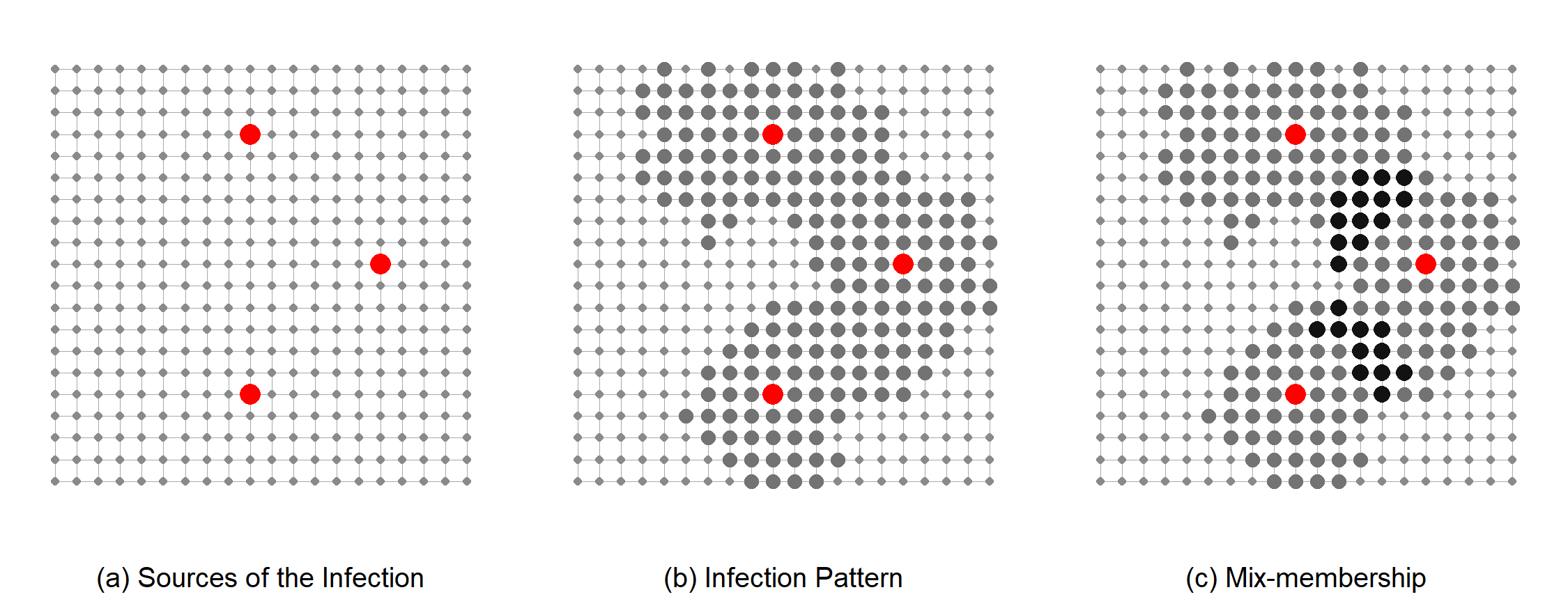}
\caption{{\bf Visualization of a MSD problem on a grid graph.}
The left panel identifies the sources of infection (red nodes), while the middle panel shows the resulting infection pattern, with enlarged nodes representing infected individuals forming clusters around each source. The right panel demonstrates the mixed-membership issue, where black nodes are influenced by multiple sources. }
\label{fig1}
\end{figure}

A significant limitation of current clustering-based MSD algorithms is their failure to account for the mixed-membership issue caused by the diffusion process. (Mixed-membership refers to the ubiquitous situation in which nodes belong to more than one cluster.) Fig \ref{fig1} illustrates the application of a clustering algorithm for the MSD problem. In the left plot, red nodes represent the sources of infection. The middle plot shows the infection pattern after a certain period, where the enlarged nodes denote infected individuals.
Typically, the infection originates from the sources and spreads locally, affecting nearby nodes and forming distinct clusters or communities centered around each source. However, as shown in the right plot, black nodes represent infected individuals influenced by more than one source. This demonstrates the mixed-membership problem, where nodes belong to multiple communities. Ignoring this issue can lead to overestimating the number of clusters in the detection process.

%  \textcolor{red}{Um, forgetting something?}
Edge clustering algorithms offer a powerful solution to this challenge by clustering edges rather than nodes, thereby capturing relationships between nodes and accounting for scenarios where nodes belong to multiple sources or communities. Key algorithms in this domain include the Latent Space Edge Clustering (LSEC) model~\cite{sewell2021model}, which clusters edges based on latent features of nodes; and the automated Latent Space Edge Clustering (aLSEC) model~\cite{pham2024automated}, which builds on LSEC by automating the determination of the number of clusters, which incorporates edge weights and accounts for noisy or anomalous edges. These models represent significant advancements in edge clustering by addressing limitations in traditional node-based methods, particularly in handling mixed-membership.

Building on these developments, this paper integrates the edge clustering framework with existing multiple-source detection solutions, specifically the Community-based Label Propagation (CLP) framework. By leveraging the ability of edge clustering to address mixed-membership, the proposed approach enhances the precision and accuracy of source detection.

In Section \nameref{algorithm-4}, we detail the entire process of our proposed algorithm. Section \nameref{simulation4} presents the results of a simulation study designed to evaluate the effectiveness of this method. In Section \nameref{result-4}, we summarize the outcomes of the simulations, while Section \nameref{discussion4} provides a discussion of the key findings and their implications.

\section*{Algorithm}
\label{algorithm-4}
Our proposed algorithm builds upon the CLP framework, extending it to account for the possibility that nodes may belong to multiple sources. We begin with the full network, denoted as $G = (\mathcal{V}, \mathcal{E})$, where $\mathcal{V}$ is the set of nodes and $\mathcal{E} \subset \mathcal{V}\times \mathcal{V}$ is the set of edges between elements of $\mathcal{V}$. Given the full network $G$, the observed infected sub-network is represented as $G_I := (\mathcal{V}_I, \mathcal{E}_I)$, where $\mathcal{V}_I \subseteq \mathcal{V}$ and $\mathcal{E}_I = \mathcal{E}\cap(\mathcal{V}_I\times\mathcal{V}_I).$ In this scenario, it is assumed that all infected or influenced nodes are fully observed. We additionally consider information about uninfected boundary nodes, denoted as $\mathcal{V}_{UB}$, which are adjacent to at least one infected neighbor. Together, the infected sub-network and the boundary nodes form the extended infected network, $G_{EI} = (\mathcal{V}_{EI}, \mathcal{E}_{EI})$, where $\mathcal{V}_{EI} = \mathcal{V}_{UB} \cup \mathcal{V}_I$ and $\mathcal{E}_{EI} = \mathcal{E}\cap(\mathcal{V}_{EI}\times\mathcal{V}_{EI}).$ The objective of the algorithm is to identify the original infection source set, $S \subset \mathcal{V}_I$.

\subsection*{Edge Clustering}
Let $G_I$ consist of $N_I$ nodes and $M_I$ edges. The edge set $\mathcal{E}_I = \{ e_m \}_{m=1}^{M_I}$ represents the edges in $G_I$, where each edge is defined as $e_m = (e_{m1}, e_{m2})$, with $e_{m1}$ being the sending node and $e_{m2}$ the receiving node. We apply an edge clustering algorithm on $G_I$, detecting $K$ distinct clusters. Rather than prespecifying $K$, by using the aLSEC algorithm $K$ is automatically determined in a data-driven manner. See \cite{pham2024automated} for more details. Let $Z_{mk}\in\{0,1\}$ indicate whether edge $e_m$ belongs to cluster $k$.

\subsection*{Label Assignment}
For any infected node $u$, the node's age is defined as in \cite{ali2019epa}:
\begin{equation*}
    A_u^I = \frac{I_u}{O_u}(1+\log O_u)
\end{equation*}
where $I_u$ represents the number of infected neighbors of node $u$ (i.e., the degree of node $u$ in the infection network $G_I$) and $O_u$ is node $u$'s degree in the original network $G$.  
For any uninfected node $v$ in $G_{EI}$, the age is defined as:
\begin{equation*}
    A_v^U = \frac{\sum_{u \in N_I(v)} I_u}{I_v}
\end{equation*}
where $N_I(v)$ denotes the set of infected neighbors of node $v$, and as before $I_u = |N_I(u)|$ is the number of infected neighbors.

The concept of node age for both infected and uninfected nodes is introduced to evaluate their prominence and incorporate exoneration effects. The prominence effect suggests that nodes surrounded by a larger proportion of infected neighbors are more likely to be identified as the infection sources, as their prominence in the network makes them strong candidates for initiating the spread. Conversely, the exoneration effect posits that neighboring nodes can act like alibis, reducing the likelihood of a node being the source. For example, if a node is infected but surrounded by a high proportion of uninfected neighbors, this lowers the probability of that node being the true source, as the uninfected neighbors provide evidence against the node being responsible for the spread \cite{ali2019epa}.

Following the edge clustering in the first step, $K$ clusters are obtained for the infected network $G_I$. We then consider the extended infected network $G_{EI}$, which includes $N_{EI}$ nodes ($N_{EI} \geq N_I$). We construct a $N_{EI} \times (K+1)$ label matrix $\mathcal{L}^t$, where $t$ represents the iteration number. Initially, we set $\mathcal{L}^0$ using the results from the edge clustering in the following way. For $k=1,\ldots,K$, set
\begin{align} \nonumber
	\mathcal{L}^{0}_{ik} 
	&:= 
	\begin{cases}
		A_i^I & \mbox{if $i\in \mathcal{V}_I$ and } \exists m:Z_{mk}=1 \text{ \& }e_{m1}=i\cup e_{m2}=i \\
		0 & \mbox{otherwise},
	\end{cases}
%    \mathcal{L}^{0}_{ik} 
%    &:= 
%    \begin{cases}
%        A_i^I & \mbox{if $i\in \mathcal{V}_I$ and } \exists m:e_{m1}=i\cup e_{m2}=i \\
%        0 & \mbox{otherwise},
%    \end{cases}
    & \\
    \mbox{and }\mathcal{L}^{0}_{i(K+1)} 
    &:= 
    \begin{cases}
        \underset{v\in \mathcal{V}_{UB}}{\max}\big(A_U(v)\big) - A_U(i) & \mbox{if $i\in \mathcal{V}_{UB}$} \\
        0 & \mbox{otherwise}
    \end{cases} 
\end{align}
% At the beginning, $\mathcal{L}^0$ is initialized to all zero. Denote each edge $e_m = (e_{m1} = i, e_{m2} = j)$ in $G_I$, where $e_{m1} = i$ denotes the sending node, $e_{m2} = j$ denotes the receiving node, and $w_m$ is the edge weight. If $Z_{mk} = 1$, we assign $\mathcal{L}^0[i, k] = A_i^I$ and $\mathcal{L}^0[j, k] = A_j^I$. This process populates the first $K$ columns of $\mathcal{L}^0$. 
% 
% The $(K+1)^{\text{th}}$ column captures the exoneration effect. For any uninfected node $i$ where $i \in \mathcal{V}_{UB}$, we assign $\mathcal{L}^0[K+1, i] = \text{MaxAge} - A_U(i)$, where $\text{MaxAge}$ represents the maximum age among all uninfected nodes in $G_{EI}$. 
The reasoning behind this assignment is that older uninfected nodes exert a weaker exoneration effect on their infected neighbors, as their prominence reduces with increasing age.

\subsection*{Label Propagation}  

The label matrix $\mathcal{L}$ is propagated iteratively throughout the network. In each iteration, nodes in the extended infected network %\textcolor{red}{It looks to me from the equation that all nodes, not just the infected nodes, get updated.  Can we remove the word ``infected'' from this sentence?} 
update their labels by incorporating both the prominence and exoneration effects from their neighboring nodes. Nodes with higher prominence are more likely to propagate stronger labels (ages) to their neighbors.

\begin{equation*}
    \mathcal{L}^{t+1} = \alpha \mathcal{A} \mathcal{L}^t + (1-\alpha) \mathcal{L}^0
\end{equation*}

where $\mathcal{A}$ represents the symmetrically normalized adjacency matrix of the extended infection graph $G_{EI}$, and $0 < \alpha < 1$ is the weight that determines the proportion of label information each node receives from its neighbors versus its initial value. In this study we set $\alpha$ to be $0.5$.

It has been proven that the label matrix $\mathcal{L}^t$ will converge to a stationary value: \cite{wang2017multiple}

\begin{equation*}
    \mathcal{L}^* = (1 - \alpha)(I - \alpha \mathcal{A})^{-1}\mathcal{L}^0
\end{equation*}

\subsection*{Source Identification} 

After obtaining the converged label matrix $\mathcal{L}^*$, we then row-normalize $\mathcal{L}^*$ to obtain node-specific cluster source scores. The node in $\mathcal{V}_I$ with the highest score in each of the first $K$ columns of the row-normalized $\mathcal{L}^*$ is selected as the infection source for the corresponding community. Mathematically, the source node for community $k$, denoted as $S_k$, is given by:

\begin{equation*}
S_k = \arg \max_{i \in \mathcal{V}_{I}} \mathcal{L}^*[i,k] \quad \text{for} \quad k = 1, 2, \dots, K
\end{equation*}

\section*{Simulation Design}
\label{simulation4}

The simulation study was built on the real-world datasets. We selected three networks from Adolescent Health, or ADD HEALTH Networks \cite{moody2001peer} in the networkdata R package \cite{networkpackage}. The ADD HEALTH data is based on a 1994-95 survey where 90,118 students from 84 communities participated, representing school-based social networks. In some communities, there were two schools, allowing students to name friends from a ``sister school." Each student filled out a questionnaire, naming up to five male and five female friends.

We selected three networks from the ADD HEALTH dataset: addhealth15, addhealth20, and addhealth75. Table \ref{tab:sim_data} summarizes the features of these networks. At the start of each simulation, $K$ sources were randomly selected from each network, and the infection process was simulated with an infection probability of 0.2. The spread continued until over $10\%$ of the nodes in the network were infected. Different clustering algorithms were then applied within the CLP framework. Specifically, we tested three methods: (1) Louvain, (2) Leading Eigenvector, and (3) aLSEC. Louvain clustering and the Leading Eigenvector method are widely recognized as state-of-the-art node clustering algorithms for addressing clustering-based source detection problems \cite{zhang2023clp, zang2015locating}.

\begin{table}[!ht]
\centering
\centering
\caption{
{\bf The Description of Datasets.}}
\begin{tabular}{l|cccc}
Network    & \multicolumn{1}{l}{\# Nodes} & \multicolumn{1}{l}{\# Edges} & \multicolumn{1}{l}{Ave. degree} & \multicolumn{1}{l}{Density} \\ \hline
addhealth15 & 1089     & 5370     & 9.86         & 0.0045      \\
addhealth20 & 922      & 5229       & 11.34       & 0.0062 
\\
addhealth75 & 1011      & 5459       & 10.80       & 0.0053
\end{tabular}
\label{tab:sim_data}
\end{table}

% The decision to use the aLSEC model over the WLSEC model stems from the nature of the ADD HEALTH networks, which have discrete edge weights ranging from 1 to 6. Since the WLSEC model currently assumes that edge weights follow a normal distribution, it is not suitable for handling such discrete values. As a result, the ADD HEALTH networks were converted into unweighted networks to ensure compatibility with the aLSEC model.

For each network, we set the number of sources 
$K$ to 1, 3, and 5. Under each configuration, we conducted 200 simulations with varying random seeds. The accuracy of each source detection approach was evaluated using the F1-Measure \cite{van1979information}, which combines precision and recall to provide an overall performance metric. %The F1-Measure is calculated as the harmonic mean of precision and recall, defined as:
%\begin{equation}
%    \text{F1-Measure} = \frac{2 \times precision \times recall}{precison + recall}
%    \label{eq:fmeasure}
%\end{equation}
%
%Here, precision represents the proportion of correctly identified sources out of all retrieved sources, as defined in Eq: \ref{eq:precison}, and recall refers to the proportion of correctly identified sources relative to the total number of actual sources, as shown in Eq: \ref{eq:recall}.
%
%\begin{equation}
%    precision = \frac{ \{ \text{retrived sources}  \} \cap \{ \text{true sources}  \} }{\{ \text{retrived soures}  \}}
%    \label{eq:precison}
%\end{equation}
%
%\begin{equation}
%    recall = \frac{ \{ \text{retrived sources}  \} \cap \{ \text{true sources}  \} }{\{ \text{true soures}  \}}
%    \label{eq:recall}
%\end{equation}

Computation for the simulation study was done on University of Iowa High-performance Computing (HPC) system. The tests of model computation time were performed on a server with an Intel(R) Xeon(R) Gold 6230 CPU @ 2.10 GHz. All code was executed by R in version 4.0.5 \cite{Ritself2023}. 

\section*{Result}
\label{result-4}

\begin{figure}[!h]
\centering
  \includegraphics[scale=0.45]{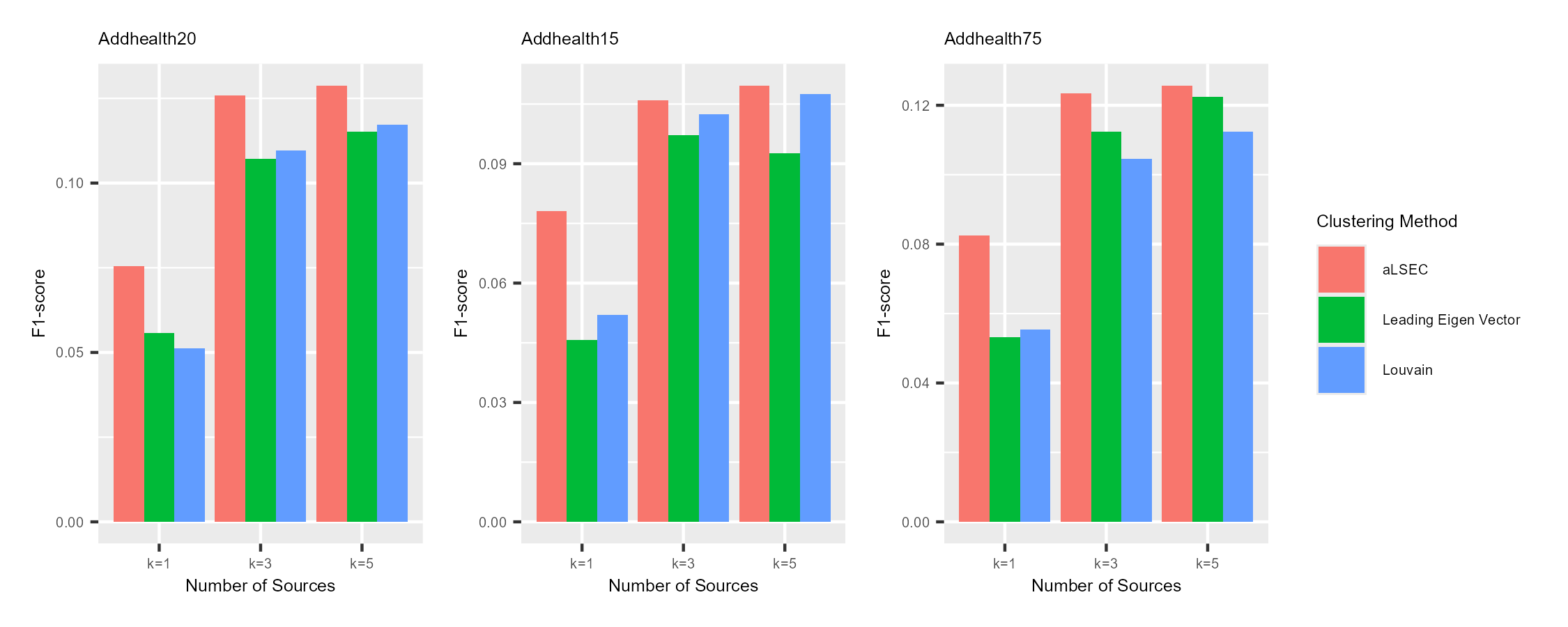}
\caption{{\bf Simulation Results.}
Comparison of F1-Measure performance across clustering methods for multiple-source detection (MSD) in three ADD HEALTH networks: Addhealth20, Addhealth15, and Addhealth75. Each panel corresponds to a different network. }
\label{fig2}
\end{figure}

The simulation results presented in Figure \ref{fig2} demonstrate that the aLSEC model consistently outperforms both the Louvain and Leading Eigenvector clustering algorithms across all three selected ADD HEALTH networks (addhealth15, addhealth20, and addhealth75). The y-axis represents the F1-Measure, which evaluates the accuracy of source detection, while the x-axis shows the number of sources tested (\(K=1\), \(K=3\), and \(K=5\)).

% \textcolor{red}{Again, forgetting something?}

In all cases, the aLSEC model achieves the highest F1-Measure scores, indicating its superior performance in accurately detecting the true sources of infection or diffusion in these networks. This trend holds across both single-source and multiple-source detection scenarios, with the difference in performance being particularly pronounced when \(K=1\). This suggests that the aLSEC model is especially effective in smaller-scale infection detection settings, where it avoids the common pitfall of overestimating the number of clusters, a limitation that both the Louvain and Leading Eigenvector methods encounter due to their failure to account for the mixed-membership issue.

% Results and Discussion can be combined.
\section*{Discussion}
\label{discussion4}
This work utilizes edge clustering algorithms to address the multiple source detection (MSD) problem, focusing on the challenges of overlapping communities and mixed-membership in diffusion processes. By incorporating edge clustering into the CLP framework, we enhance traditional community-based detection methods to effectively handle scenarios where nodes are influenced by multiple sources. Simulation results demonstrate the superiority of this approach, particularly in achieving greater accuracy in multiple-source detection compared to conventional methods such as Louvain and Leading Eigenvector.

However, certain limitations remain. Firstly, this study focused on unweighted networks, utilizing the aLSEC model for edge clustering. While effective, many real-world networks involve weighted edges, where connection strengths vary across relationships. A promising extension would be incorporating the WECAN model \cite{li2025model}, an edge clustering algorithm designed for weighted networks, which builds on aLSEC by directly accounting for edge weights. Incorporating edge weights could enhance source detection accuracy by distinguishing between stronger and weaker connections. 

Additionally, this study relies on a "snapshot" observation of the network, capturing the state of infection at a single point in time. This static approach may limit the ability to fully capture the dynamics of diffusion processes. Extending the methodology to dynamic networks, where the temporal evolution of the spread is accounted for, could lead to more accurate and timely identification of sources by incorporating changes in the network structure and infection status over time. Addressing these limitations will enhance the robustness and applicability of the edge clustering algorithm for multiple-source detection, further solidifying its potential as a versatile tool for diverse network-based applications.

\section*{Acknowledgments}
This work was supported by the US Centers for Disease Control and Prevention (5 U01CK000594-04-00) as part of the MInD-Healthcare Program (\url{https://www.cdc.gov/healthcare-associated-infections/php/research/mind-healthcare.html}). %who did not play any role in the study design, data collection and analysis, decision to publish, or preparation of the manuscript.

\section*{Data availability}
Data used in our simulation study are publicly available through the R package \texttt{networkdata} \cite{networkpackage}.

% Either type in your references using
% \begin{thebibliography}{}
% Text
% \end{thebibliography}
%
% or
%
% Compile your BiBTeX database using our plos2015.bst
% style file and paste the contents of your .bbl file
% here. See http://journals.plos.org/plosone/s/latex for 

\end{document}